\documentclass{ifacconf}

\usepackage{amssymb}
\usepackage{amsmath}

\usepackage{amssymb}
\usepackage{amsmath}
\usepackage{color,soul}
\usepackage{bbm}
\usepackage{graphicx}     
\usepackage{nicematrix}
\usepackage{natbib}
\usepackage{adjustbox}
\usepackage{multirow}
\usepackage{array}
\usepackage{booktabs}
\usepackage{mathtools}
\usepackage{multicol}
\usepackage{algpseudocode}
\usepackage{algorithm}
\usepackage{xcolor}
\usepackage{caption}

\usepackage{graphicx}      
\usepackage{natbib}        

    \newcommand{\argmin}{\operatornamewithlimits{argmin}}
\begin{document}
\begin{frontmatter}

\title{Reinforcement Learning-based Control via Y-wise Affine Neural Networks: Comparative Case Studies for Chemical Processes}

\author[WVU]{Austin Braniff} 
\author[WVU]{Yuhe Tian} 

\address[WVU]{Department of Chemical and Biomedical Engineering, West Virginia University, 
   Morgantown, WV 26506 USA (e-mail: \{austin.braniff, yuhe.tian\}@ mail.wvu.edu).}

\begin{abstract}              
In this work we present an efficient and practically implementable approach for the application of reinforcement learning (RL)-based control in chemical process systems. This is an area that has yet to widely adopt RL-based control largely due to inherent challenges in trusting RL algorithms and the time-consuming process of training reliable agents. To address these challenges, we leverage a class of RL algorithms termed Y-wise Affine Neural Network (YANN)-RL, which we have developed in our prior work \citep{braniffReinforcementLearningbasedControl2025b}. By strategically initializing actor and critic networks YANN-RL algorithms provide confident and interpretable starting points within control schemes. We apply this RL-based control approach to three different process engineering case studies publicly available on the PC-Gym library \citep{bloorPCGymBenchmarkEnvironments2026}: (i) a continuous stirred tank reactor (CSTR), (ii) a four-tank system, and (iii) a multistage extraction column. Our approach is compared to several popular RL algorithms (PPO, SAC, DDPG, and TD3) and is benchmarked against nonlinear model predictive control (NMPC). These case studies demonstrate that YANN-RL can greatly reduce the training time and data needed, can be deployed with confidence for chemical process systems, and can approach the performance of NMPC without the knowledge of a full nonlinear model. 
\end{abstract}

\thanks{© 2026 the authors. This work has been accepted to IFAC for publication under a Creative Commons Licence CC-BY-NC-ND.}


\end{frontmatter}
\section{Introduction}
Reinforcement learning (RL) has gained significant attention in recent years, with early breakthroughs in game playing demonstrating the power of combining neural networks with RL algorithms \citep{mnihHumanlevelControlDeep2015c}. Although RL has been applied successfully across several fields such as robotics, autonomous decision-making, process design, etc., its relevance to the control of chemical process systems has been distinctly lacking \citep{devarakondaRecentAdvancesReinforcement2025}. RL has been investigated for a few examples of process control applications including distillation, chemical reactors, and multi-tank systems \citep{spielbergSelfdrivingProcessesDeep2019,fariaOneLayerRealTimeOptimization2023,dogruOnlineReinforcementLearning2021}, where it has shown strong performance.

Despite these successes, RL has not yet achieved prevalent adoption in chemical process control. The primary barriers are concerns over safety and interpretability, particularly in the early exploration phases \citep{dogruReinforcementLearningProcess2024,shinReinforcementLearningOverview2019a}. During exploration, RL agents often test unproven or random actions to improve performance, which can lead to unsafe behavior. This is an unacceptable risk in industrial settings \citep{nian2020review}. This inherent non-interpretability and random initializations of policies within well-established RL algorithms pose several challenges. Additionally, RL algorithms are known for being data-hungry and may require exceptionally long training times to achieve acceptable control for complex systems. These issues hinder the deployment of RL-based controllers in safety-critical process environments that demand reliability, safety, and clear operational insight \citep{fariaWhereReinforcementLearning2022}.

To address these challenges, significant efforts have been made in recent years to improve the trust and practicality of RL algorithms. One strategy is to pre-train an RL policy network using data generated by more trusted control methods, such as model predictive control (MPC). These strategies preserve a model-free RL framework, linear MPC can be used, as linear system models are relatively easy to approximate through techniques like system identification \citep{hassanpourPracticalReinforcementLearning2025}. Towards more algorithmic improvements on deploying RL with confidence, recent approaches have investigated the use of Lyapunov functions within an RL framework. Lyapunov functions, Lyapunov surrogates, or control Lyapunov barrier functions are used to guarantee certain mathematical properties of the RL-based controller \citep{changStabilizingNeuralControl2021,wangControlLyapunovbarrierFunctionbased2024}. Similarly, control invariant sets (CIS) can also be used to satisfy certain properties of the control actions if they are known a priori \citep{boControlInvariantSet2023}. While these techniques do provide significantly more trust in RL, they rely on detailed system insight, accurate models, and can require significant computational power for systems with high degrees of nonlinearity. They also require an exploration phase during system modeling or RL agent training, which can pose significant safety risks in practical applications. This highlights the need for an RL-based control algorithm that enhances interpretability, confidence, and computational efficiency while eliminating unsafe exploration. To this end, we have developed Y-wise Affine Neural Network (YANN)-based RL in our prior work \cite{braniffReinforcementLearningbasedControl2025b}, however this work was limited to only one process control example and lacked convincing benchmarking against other RL algorithms.

In this work, we present a series of comparative case studies that investigate the use of RL-based control on three nonlinear chemical process systems: a CSTR, a four-tank system, and a multistage extraction column. We show that YANN-RL is substantially more efficient than other common RL algorithms and can achieve superior performance with significantly fewer update steps. Additionally, this paper highlights how YANN-RL can be implemented with significantly more confidence and trust due to its initialization techniques rooted in optimal control theory. Furthermore, we show how these algorithms can approach the performance of nonlinear model predictive control (NMPC) via comparison. 

The remaining sections of this work are organized as follows: Section 2 gives a brief introduction to actor-critic reinforcement learning. Section 3 covers the fundamentals of YANN-RL. Section 4 presents the comparative case studies on chemical process systems. Section 5 provides concluding remarks and discusses opportunities for future research.
\section{Actor-Critic Reinforcement Learning}
This work focuses on a class of RL algorithms termed actor-critic algorithms, we will introduce the concepts of actor-critic RL and the nomenclature from a traditional control theoretic perspective. Typically, in RL there are three main components: (i) the environment, which exists in a state, $x_t$, that takes an action, $u_t$, and transitions to a new state $s_{t+1}$, (ii) an agent that takes the current state of the environment and provides a control action by representing a control policy $\pi(s)$, and (iii) a well-designed cost function, $C(s,u)$, which is representative of the desired system behavior (e.g., the cost is low when the system is performing desirably). RL algorithms iteratively update the actor to provide better control actions by minimizing the cost function. In actor-critic RL, the agent is instead termed an actor and maintains the functionality of being the control policy. A new component is added called the critic, which represents the state-action value function (or Q-function), $Q(s,u)$, of the environment. This is conceptualized in Fig. \ref{fig:Actor-Critic RL} below. 

\begin{figure}[ht]
    \centering
    \includegraphics[width=\linewidth]{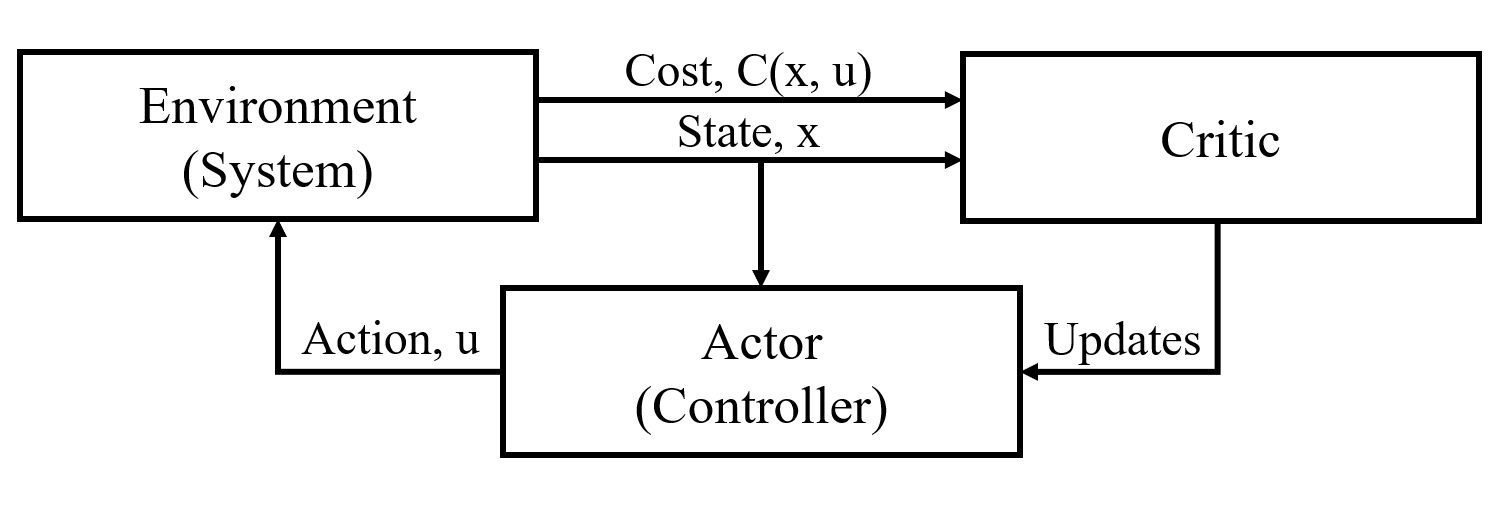}
    \caption{Actor-critic reinforcement learning with nomenclature for this work.}
    \label{fig:Actor-Critic RL}
\end{figure}

The Q-function, given in Eq. \ref{eq:Q_fun}, can be interpreted as an indicator of how desirable it is for the environment to exist in a particular state, $x_0$, while taking a particular action, $a_0$ and then follow a theoretically optimal control policy for all time after. The critic is updated iteratively to better approximate this function based on the Bellman recursive optimality principle, which is shown for the Q-function in Eq. \ref{eq:Q_opt}. 
\begin{equation}
\label{eq:Q_fun}
    Q(x_0,u_0) = C(x_0,u_0) + \sum_{t=t_1}^\infty \gamma^t C(x_t, u_t^*)
\end{equation}
\begin{equation}
    \label{eq:Q_opt}
    Q^*(x_t,u_t) = C(x_t,u_t) + \min_{u_{t+1}}\gamma Q^*(x_{t+1},u_{t+1})
\end{equation}
The actor is updated to minimize the Q-estimate given by the critic which in turn trains it to provide better control actions. If the actor is parameterized by a function with decision variables $\theta_\pi$ and the critic is parameterized by a function with decision variables, $\theta_Q$ then their update steps can be expressed by Eq. \ref{eq:actor_update} and Eq. \ref{eq:critic_update}.  
\begin{equation}
\label{eq:actor_update}
    \min_{\theta_\pi} Q(x_t,u_t)
\end{equation}
\begin{equation}
\label{eq:critic_update}
    \min_{\theta_Q} (Q(x_t,u_t)- (C(x_t,u_t) + Q(x_{t+1},u_{t+1}) )^2
\end{equation}

\section{Y-wise Affine Neural Networks in Reinforcement Learning}
\subsection{Y-wise Affine Neural Networks}

Y-wise affine neural networks (YANNs) are a neural network architecture introduced in our earlier work \citep{braniffYANNsYwiseAffine2025a}. They can exactly represent piecewise-affine functions of arbitrary input and output dimensions, defined over any number of subdomains. Piecewise affine functions are the form of the explicit solutions to linear MPC or multi-parametric MPC (mp-MPC) problems. In this way, YANNs can encode the explicit control solution produced by solving mp-MPC. When used this way, they act as NN-based controllers that retain mp-MPC’s guarantees of stability and recursive feasibility for linear systems. A formal proof of this behavior and a construction procedure for YANN's weights and biases is provided in  \cite{braniffYANNsYwiseAffine2025a}. The functionality of YANNs for a constrained regulation control problem is conceptualized by Eq. \ref{eq:YANN_MPC}.
\begin{align}
\label{eq:YANN_MPC}
YANN(x_t) &= \argmin_u  \sum_{t=0}^{N-1} \left( x_t^T Q_w x_t + u_t^T R u_t \right) + x_N^T P x_N \nonumber \\
\text{s.t.} \quad & x_{t+1} = A x_t + B u_t, \quad \forall t \in \{0,1, 2, \cdots, N-1\} \nonumber \\
& x_t \in \mathcal{X}, u_t \in \mathcal{U}, \quad \forall t \in \{0,1, 2, \cdots, N-1\} \nonumber \\
& x_{N-1} \in \mathcal{X}_f
\end{align}

It is important to note that YANNs are an offline representation of the solution to this problem and do not require anything to be resolved online other than a neural network evaluation. YANNs offer a compelling tool for RL by enabling policy networks to be initialized with a control law that is theoretically reliable, at least with respect to a linear model. 

\subsection{YANN-based Actors}

\begin{figure}[h]
    \centering
    \includegraphics[width=0.85\linewidth]{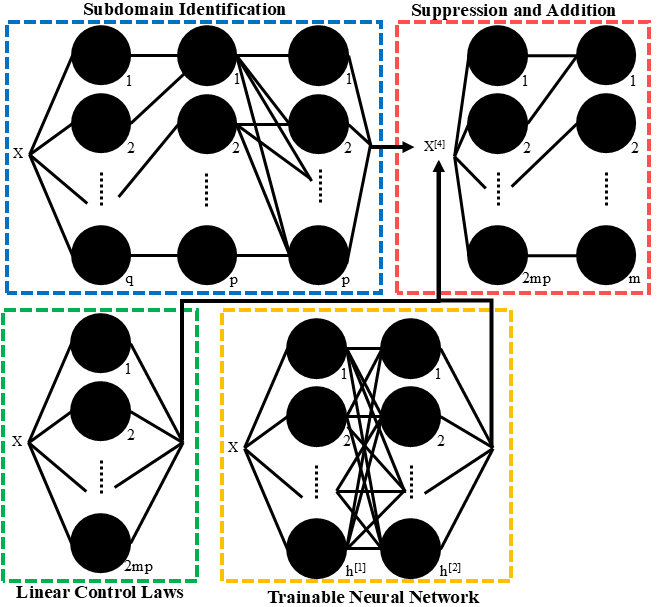}
    \caption{Conceptualization of YANN-Actor}
    \label{fig:YANN-actor}
\end{figure}

The YANN-Actor (Fig. \ref{fig:YANN-actor}) follows the YANN framework to directly represent known optimal control laws. To create a YANN-initialized policy network, a linear MPC problem is first solved offline to obtain a piecewise-affine explicit controller, typically using a simplified linear model derived from process data or some other system identification technique. The YANN can then be constructed following the procedure in \cite{braniffYANNsYwiseAffine2025a}, with additional layers that are embedded into the structure of a YANN in a particular way such that it allows the YANN-Actor to be as expressive as any other neural network without sacrificing its initial control policy. In this way, the YANN-actor can be trained similarly to an actor network from any RL algorithm, but it is guaranteed to initially be functionally equivalent to a linear optimal control policy. 

\subsection{YANN-based Critics}
\begin{figure}[ht]
    \centering
    \includegraphics[width=0.85\linewidth]{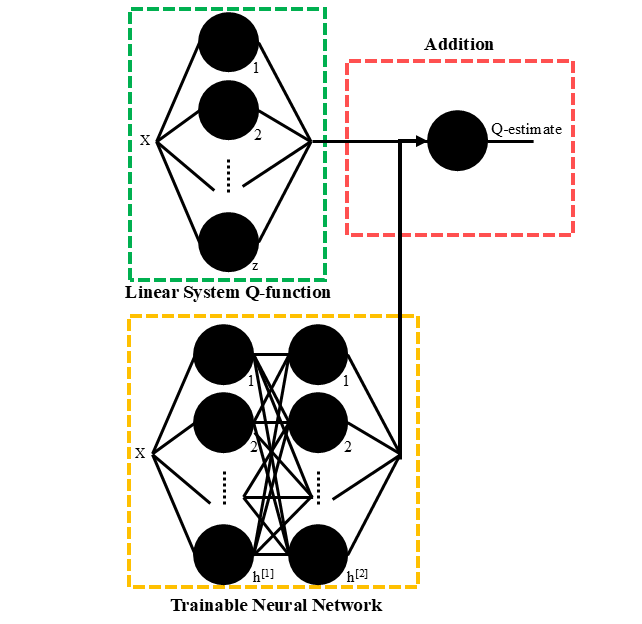}
    \caption{Conceptualization of YANN-Critic}
    \label{fig:YANN-critic}
\end{figure}

The YANN-Critic (Fig. \ref{fig:YANN-critic} stems from a separate concept than the YANN since the critic within an RL algorithm needs to represent a separate function altogether. The nomenclature is maintained for simplicity. To develop the YANN-critic, the explicit Q-function for the linear system needs to be solved for. This function is not widely used in traditional control, but as discussed previously, it is necessary in actor-critic RL. In general for an unconstrained linear system, the state-action value function can be derived as follows:
\begin{align*}
    Q^\star(s_t,u_t)& =  C(s_t,u_t) + \min_{u_{t+1}}\gamma Q^\star(s_{t+1},u_{t+1})\\
    \min_{u_{t+1}}\gamma Q^\star(s_{t+1},u_{t+1})& =  \gamma V^\star(s_{t+1}) = s_{t+1}^TPs_{t+1}\\
    Q^\star(s_t,u_t)& =  s_t^TQ_ws_t+u_t^TRu_t + \gamma s_{t+1}^TPs_{t+1}\\
    s_{t+1} &=  As_t+Bu_t\\
    Q^\star(s_t,u_t) &=  s_t^TQ_ws_t+u_t^TRu_t\\
    &+\gamma (As_t+Bu_t)^TP(As_t+Bu_t)\\
\end{align*}

Similarly to how the YANN embeds control laws, the resulting Q-function for the linear system approximation can be exactly represented within a neural network. The resulting neural network only has the capability to represent functions of similar forms if trained. To address this issue, additional network layers can be embedded into this structure in order to provide general nonlinear expressibility. In this way, the YANN-Critic, initially, is functionally equivalent to the explicit state-action value function for a linear surrogate while being as trainable as any other critic network. Further details on building YANN-Critics can be found in \cite{braniffReinforcementLearningbasedControl2025b}.

\section{Case studies}

All case studies for evaluating YANN-RL utilize the YANN-DDPG algorithm, which is summarized in Algorithm 1. In all figures, Oracle refers to well-tuned NMPC which assumes perfect and noiseless nonlinear models as an ideal benchmark. Control performance metrics such as integral squared error (ISE), integral time-weighted absolute error (ITAE), steady-state error ($e_{SS}$), and cumulative cost are reported within each study.

\begin{algorithm}
\label{alg:YANN-DDPG}
    \caption{YANN-DDPG Summary}
    \begin{algorithmic}[1]
        \State Obtain a linearized model for the environment (e.g., Jacobian linearization, system identification, etc.)
        \State Use the linear model to create an MPC problem
        \State Reformulate the MPC problem into an mp-MPC problem and solve to obtain the piecewise-affine control law
        \State Use the piecewise affine control law to initialize a YANN-actor, $\pi_{\theta}(s_t)$
        \State Use the system model and MPC problem weighting matrices to initialize a YANN-critic, $Q_{\phi}(s_t,u_t)$
        \State Initialize target networks $\pi_{\theta_{target}}(s_t)$ and $Q_{\phi_{target}}(s_t,u_t)$ with $\theta_{target} \gets \theta$, $\phi_{target} \gets \phi$
        \State Follow line 3 and onward from DDPG Algorithm excluding random exploration
    \end{algorithmic}
\end{algorithm}

\subsection{CSTR}
Continuous stirred tank reactors (CSTRs) are the standard benchmarking system in process control applications. The CSTR in this study is from \cite{seborg2016process}. It has two chemical species, $A$ and $B$ and considers a single exothermic reaction, $A\xrightarrow{}B$. There are two state variables for this model, the concentration of species $A$, $C_A$, and the temperature of the reactor, $T$. It has one manipulated variable which is the temperature of the cooling water used to combat the exothermic nature of the reactor, $T_C$. The ODEs describing this system are given in Eq. \ref{eq:CSTR}. Detailed modeling information can be found in Table 1 of \cite{bloorPCGymBenchmarkEnvironments2026}.

\begin{align}
\label{eq:CSTR}
  \frac{\mathrm{d}C_A}{\mathrm{d}t} &= \frac{q}{V}(C_{A_f} - C_A) - kC_Ae^{\frac{-E_A}{RT}}\\
  \frac{\mathrm{d}T}{\mathrm{d}t} &= \frac{q}{V}(T_f - T) -\frac{\Delta H_R}{\rho C_p}kC_Ae^{\frac{-E_A}{RT}} + \frac{UA}{\rho C_p V}(T_c - T)\nonumber
\end{align}

\begin{figure}[h]
    \centering
    \includegraphics[width=\linewidth]{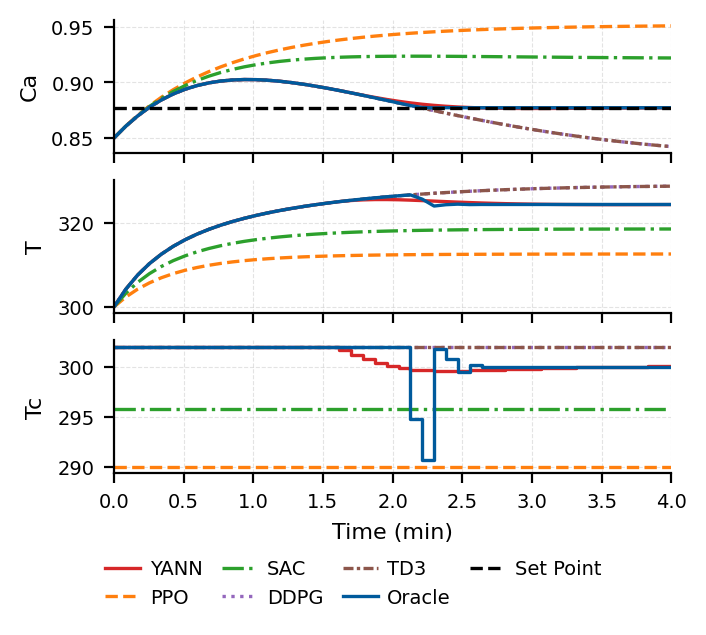}
    \caption{Control studies of RL algorithms on a CSTR.}
    \label{fig:CSTR}
\end{figure}

The objective for the operation of the CSTR is to maintain a species $A$ concentration of $C_A=0.877$ by manipulating the cooling water temperature. An initial operating point is randomly generated, and training episodes are simulated for four minutes of process time. The YANN-DDPG algorithm is trained for a total of twenty-five episodes, while all other RL algorithms are trained for an equivalent of one hundred episodes each. This gives each other RL algorithm a four times advantage in the number of training steps performed. The results of the RL training for all five RL algorithms are highlighted in Fig. \ref{fig:CSTR}. Performance metrics for this study are given in Table \ref{tab:CSTR} which shows that the YANN-RL controller attained near-perfect tracking performance by closely mimicking NMPC behavior. YANN-RL performed significantly better than all of the RL baselines despite being trained for four times fewer episodes. TD3 was the strongest RL baseline however, its tracking and steady-state errors as well as cumulative cost remained notably higher than YANN-RL.

\begin{table}[h]
    \centering
    \caption{Performance metrics for control of a CSTR.}
    \label{tab:CSTR}
    \begin{tabular}{ccccc}
    \hline
    Controller & ISE & ITAE & $e_{SS}$ & Cumulative Cost \\
    \hline
        NMPC & 0.0007 & 0.0338 & 0.0000 & 1.772\\
         YANN-DDPG & 0.0007 & 0.0378 & 0.0001 & 1.788\\
         Best RL (TD3) & 0.0015 & 0.1562 & 0.0350 & 4.396 \\
         \hline
    \end{tabular}
\end{table}

\subsection{Four-tank System}
The four-tank system is a common benchmark in process control, it features four connected tanks that exhibit strong nonlinear behavior \citep{845876}. The four-tank model consists of four states, $h_i$, corresponding to the height of the water in tank $i$ and two manipulated variables, $v_1$ and $v_2$, corresponding to the voltages applied to the pumps in the system. The equations governing this system can be found in Eq. \ref{eq:four_tank}. Detailed information regarding modeling can be found in Table 4 of \cite{bloorPCGymBenchmarkEnvironments2026}.

\begin{align}
\label{eq:four_tank}
\frac{dh_1}{dt} &= -\frac{a_1}{A_1}\sqrt{2g_ah_1} + \frac{a_3}{A_1}\sqrt{2g_ah_3} + \frac{\gamma_1 k_1}{A_1}v_1 \\
\frac{dh_2}{dt} &= -\frac{a_2}{A_2}\sqrt{2g_ah_2} + \frac{a_4}{A_2}\sqrt{2g_ah_4} + \frac{\gamma_2 k_2}{A_2}v_2\nonumber \\
\frac{dh_3}{dt} &= -\frac{a_3}{A_3}\sqrt{2g_ah_3} + \frac{(1-\gamma_2)k_2}{A_3}v_2 \nonumber\\
\frac{dh_4}{dt} &= -\frac{a_4}{A_4}\sqrt{2g_ah_4} + \frac{(1-\gamma_1)k_1}{A_4}v_1\nonumber
\end{align}

\begin{figure}[h]
    \centering
    \includegraphics[width=\linewidth]{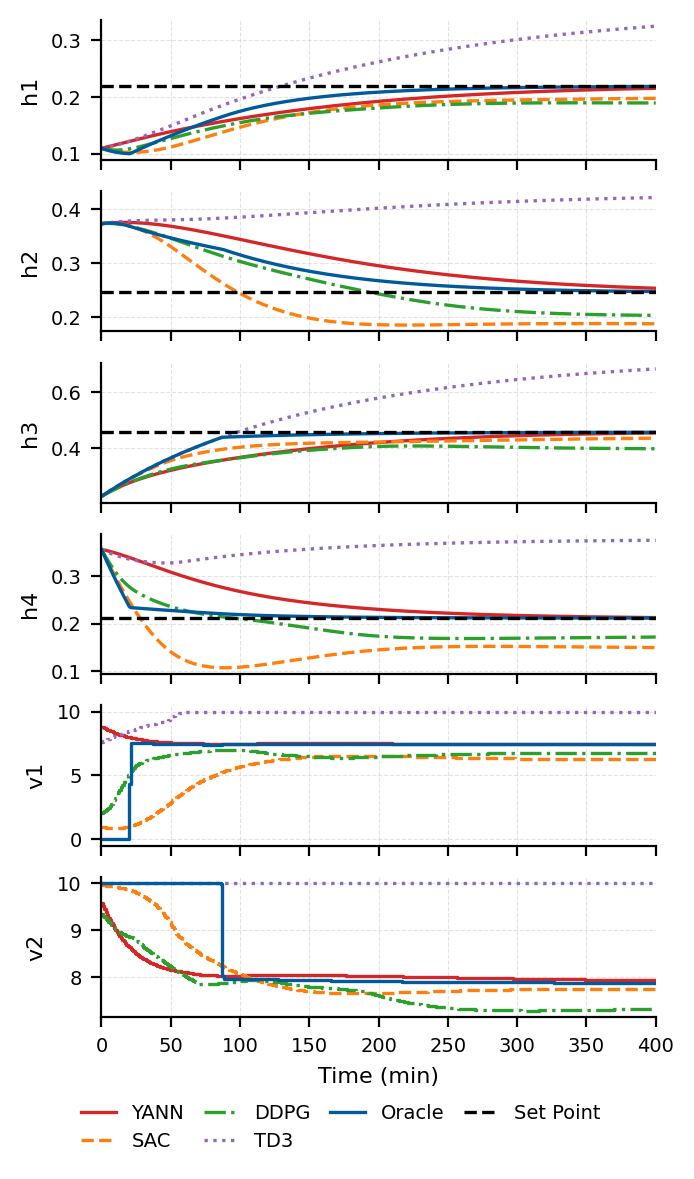}
    \caption{Control studies of RL algorithms on a four-tank system.}
    \label{fig:four_tank}
\end{figure}

For this study, the control objective is to operate around a steady state defined by $h = [0.219, 0.247, 0.457, 0.212]$. A random operating point is used for the initial state while training episodes are simulated for four-hundred minutes of process time. In this study all algorithms are trained for twenty episodes worth of updates. Due to the significantly longer simulation time, each episode corresponds to many more updates being performed per training episode. It is important to note that for some randomly generated initial states the classic RL algorithms may fail to train by driving the system to infeasible operating points. This is prevalent across all classic RL algorithms and is the reason why the PPO algorithm does not appear in the results for this system. The YANN-RL algorithm never experienced this issue. The results for the four-tank system are provided in Fig. \ref{fig:four_tank}. Control comparison metrics are provided in Table \ref{tab:four_tank} which shows that the YANN-based RL algorithm exhibits tracking and cumulative cost performance comparable to the best RL baseline (DDPG) while maintaining a significantly better steady-state error, demonstrating superior regulation control. DDPG showed a slight advantage in ISE, but the YANN-RL approach proved to be safer and more reliable by consistently avoiding infeasible operating conditions.

\begin{table}[h]
    \centering
    \caption{Performance metrics for control of a four-tank system.}
    \label{tab:four_tank}
    \begin{tabular}{ccccc}
    \hline
    Controller & ISE & ITAE & $e_{SS}$ & Cumulative Cost\\
    \hline
        NMPC & 3.775 & 2,957 & 0.0005 & 587.8 \\
         YANN-DDPG & 6.824 & 7,925 & 0.0037 & 1,075 \\
         Best RL (DDPG) & 6.507 & 13,440 & 0.0428 & 1,025\\
         \hline
    \end{tabular}
\end{table}

\subsection{Multistage Extraction Column}
The multistage extraction column is a highly complex nonlinear system with several variables governing its dynamics \citep{ingham2008chemical}. It features ten state variables, $C_{X_n}$ and $C_{Y_n}$ which represent solute liquid and gas phase concentration in each of the five stages. The column has two control inputs, the liquid flowrate, $L$, and the gas flowrate, $G$. The dynamics of this system are given in Eq. \ref{eq:column} and detailed modeling information can be found in Table. 2 of \cite{bloorPCGymBenchmarkEnvironments2026}.
\begin{align}
\label{eq:column}
\frac{dC_{X_n}}{dt} &= \frac{1}{V_l}(L(C_{X_{n-1}} - C_{X_n}) - F_n) \\
\frac{dC_{Y_n}}{dt} &= \frac{1}{V_g}(G(C_{Y_{n+1}} - C_{Y_n}) + F_n) \nonumber\\
\text{where:\quad} C_{X_{n,eq}} &= \frac{C_{Y_{n,eq}}}{m}^{e} \nonumber\\
F_n &= K_{la}(C_{X_n} - C_{X_{n,eq}})V_l\nonumber
\end{align}

The control objective for the multistage extraction column is to regulate all ten system states to a fixed operating point. This is a particularly challenging problem since all ten nonlinearly related objectives must be considered with only two manipulatable variables available. The initial operating point is randomly generated, and time profiles are simulated for 0.25 minutes. The YANN-DDPG algorithm is trained for fifty episodes, while all other RL algorithms are trained for two hundred episodes, which provides the classic RL algorithms with four times as many update steps. Trajectory tracking results for this study are given in Fig. \ref{fig:extraction column}. Performance metrics for this study are given in Table \ref{tab:extraction column} which shows that the YANN-based controller achieves performance comparable to the NMPC benchmark while substantially outperforming traditional RL methods in all control metrics. Although SAC was the best-performing RL algorithm, its error metrics and cumulative cost remained significantly higher despite being trained for four times as many episodes as YANN-RL. 

\begin{table}[h]
    \centering
    \caption{Performance metrics for control of a multistage extraction column.}
    \label{tab:extraction column}
    \begin{tabular}{ccccc}
    \hline
    Controller & ISE & ITAE & $e_{SS}$ & Cumulative Cost\\
    \hline
        NMPC & 0.0014 & 0.0005 & 0.0000 & 43.11 \\
         YANN-DDPG & 0.0015 & 0.0007& 0.0006  & 57.05 \\
         Best RL (SAC) & 0.0169 & 0.0181 & 0.0637 & 2,148\\
         \hline
    \end{tabular}
\end{table}

\subsection{Conclusion}
In this study, we demonstrated that YANN-RL offers an effective and practical pathway for bringing reinforcement learning–based control into chemical process applications, an area where RL has traditionally faced barriers related to trust and training efficiency. Across three benchmark problems: (i) a CSTR, (ii) a four-tank system, and (iii) a multistage extraction column, YANN-RL consistently outperformed several widely used RL algorithms (PPO, SAC, DDPG, and TD3) in training speed and control performance. Moreover, its performance approached that of nonlinear MPC without requiring a full nonlinear model. Overall, the results illustrate that YANN-RL provides a trustworthy, data-efficient, and high-performing RL framework suitable for real-world chemical process control. Ongoing work will be improving YANN-based RL to satisfy desirable control properties such as stability as well as introducing filtering techniques into the framework to address uncertainties. 
\begin{ack}
The authors acknowledge financial support from NSF RETRO Project CBET-2312457, NSF GRFP 2024370240, and the Department of Chemical and Biomedical Engineering at West Virginia University.
\end{ack}

\bibliography{bibliography}             
                                                   







\appendix
\section{Multistage Extraction Column Results}
\begin{figure}[h]
    \centering
    \includegraphics[width=0.89\linewidth]{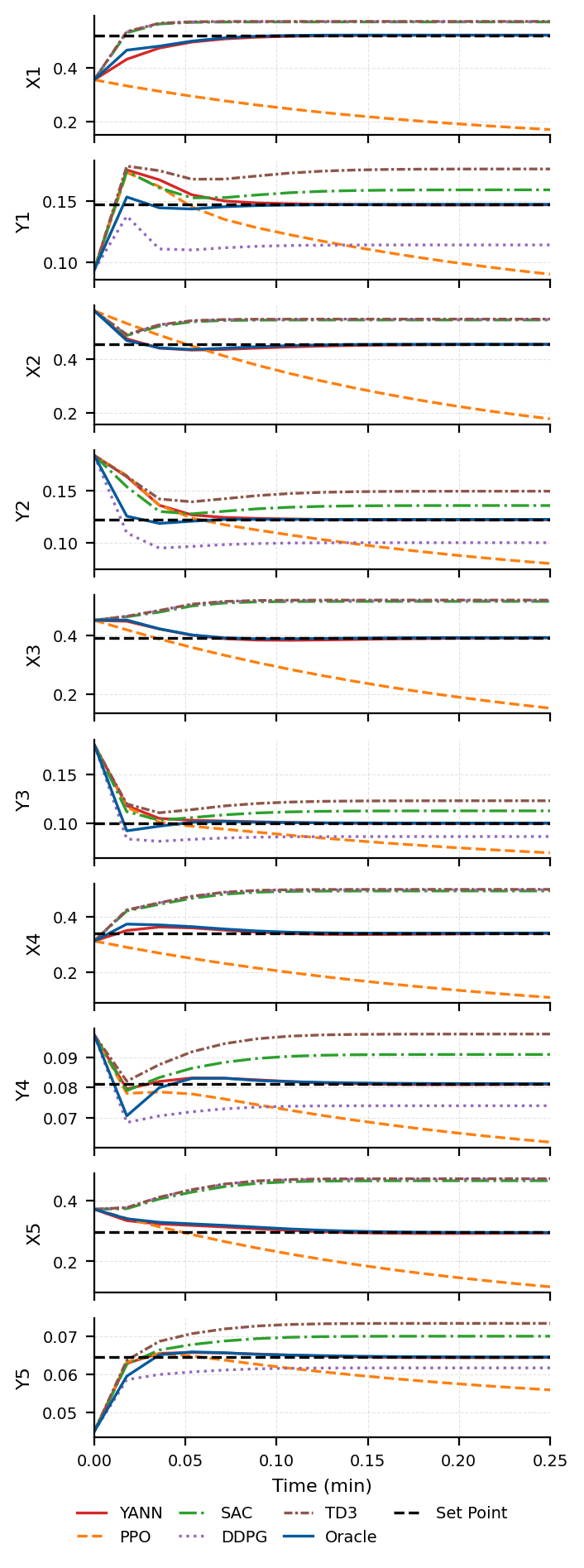}
    \caption{Control studies of RL algorithms on a multistage extraction column.}
    \label{fig:extraction column}
\end{figure}
\end{document}